\documentclass{article}

\usepackage[preprint]{neurips_2025}

\workshoptitle{Workshop on
Efficient Reasoning}
\usepackage{graphicx}

\usepackage[utf8]{inputenc} 
\usepackage[T1]{fontenc}    
\usepackage{hyperref}       
\usepackage{url}            
\usepackage{booktabs}       
\usepackage{amsfonts}       
\usepackage{nicefrac}       
\usepackage{microtype}      
\usepackage{xcolor}         
\usepackage{xspace}

\usepackage{color}
\usepackage[dvipsnames,svgnames,table]{xcolor}
\usepackage[linecolor=DarkGreen,backgroundcolor=DarkGreen!25,bordercolor=DarkGreen]{todonotes}

\newcommand{\minisection}[1]{\vspace{0.05in}\noindent {\bf #1}}

\title{Are We Scaling the Right Thing? A System Perspective on Test-Time Scaling}

\author{%
Youpeng Zhao$^{1}$\thanks{Part of the work during an internship at Microsoft Research.} \quad Jinpeng LV$^2$ \quad Di Wu$^1$ \quad Jun Wang$^1$ \quad Christopher Gooley$^2$ \\[6pt]
$^1$University of Central Florida \quad $^2$Microsoft Research
\\[6pt]
\texttt{\{youpeng.zhao,di.wu,jun.wang\}@ucf.edu} \\
\texttt{\{jinpeng.lv,gooley\}@microsoft.com} \\
}

\begin{document}

\maketitle

\begin{abstract}
Test-time scaling (TTS) has recently emerged as a promising direction to exploit the hidden reasoning capabilities of pre-trained large language models (LLMs).
However, existing scaling methods narrowly focus on the compute-optimal Pareto-frontier, ignoring the simple fact that \textit{compute-optimal is not always system-optimal}.
In this work, we propose a system-driven perspective on TTS, analyzing how reasoning models scale against practical metrics, such as latency and cost-per-token. 
By evaluating the impact of popular optimizations such as tensor parallelism and speculative decoding, our preliminary analysis reveals the limitations of current methods and calls for a paradigm shift toward holistic, system-aware evaluations that capture the true essence of scaling laws at inference time.
\end{abstract}

\section{Introduction}
The prevalence of transformer-based large language models (LLMs)~\cite{attention,gpt2} has brought the prospect of artificial general intelligence (AGI) within closer reach, transforming it from a distant dream into a practical pursuit. 
In recent years, we have witnessed a remarkable acceleration in the capabilities and accessibility of LLMs, driven by concurrent advances in model architecture~\cite{opt,llama2,deepseek}, training techniques~\cite{deepspeed,megatron}, and hardware infrastructure~\cite{cerebras,efficientscaling}. 
Powerful open-source models like LLaMA~\cite{llama2,llama}, DeepSeek~\cite{deepseek}, and Qwen~\cite{qwen2,qwen3} have democratized access to state-of-the-art language modeling, enabling a wide range of applications including intelligent assistants~\cite{cluade,chatgpt}, code generation~\cite{cursor}, and reasoning agents~\cite{chatgptagent,gemini}.
A key force behind such rapid progress has been the discovery and practice of scaling laws, which empirically show that the performance of LLMs improves predictably as a power-law function of three factors: \textit{model size}, \textit{dataset size}, and \textit{computational budget}~\cite{scaling1,scaling2}. 
These laws have provided a crucial road map for the industry, guiding the design of ever-larger models by allowing researchers and engineers to anticipate performance gains before training begins~\cite{cerebras,opt}.

However, as the LLM landscape gradually shifts from offline pre-training to online inference, these traditional scaling laws are revealing their limitations. 
\underline{First}, the dominant user cases for LLMs now center on complex reasoning tasks~\cite{gpqa,gsm8k,math500} that involve multi-step problem-solving and planning. 
Albeit larger models excel at memorization and generalization on surface-level benchmarks~\cite{openbookqa,wiki,pubmedqa}, their performance gains on tasks requiring structured reasoning can be marginal without task-specific fine-tuning~\cite{mem1,mem2,mem3}. 
Second, as LLMs quickly transition from research prototypes to production systems, inference cost, instead of training cost, becomes the dominant long-term bottleneck. 
Serving large models at scale incurs significant computational, memory, and energy overhead, particularly in latency-sensitive applications~\cite{vLLM,sglang,flexgen}. 
The interplay between model performance and the practical costs of inference represents a critical trade-off that falls outside the purview of traditional scaling laws.

To address these real-world constraints, test-time scaling (TTS) has emerged as a new promising methodology~\cite{tts1,tts2,tts3}. 
Unlike pretraining scaling laws, TTS focuses on improving model performance during the inference process. 
This paradigm encompasses powerful techniques such as supervised fine-tuning (SFT)~\cite{s1}, specialized prompting strategies~\cite{cot,tot}, and reinforcement learning~\cite{deepseekr1,gpt4}, which unlock the latent reasoning abilities of LLM at a fraction of the pre-training cost.
Despite its promise, the existing TTS paradigm remains narrowly focused on a Pareto frontier of performance vs. computation (i.e., FLOPs or the number of generated tokens). 
This ignores a critical reality of production systems: \textbf{compute-optimal does not always mean system-optimal}. 
On the one hand, previous research has shown that accelerating LLM inference demands extensive system-level optimization, such as efficient memory management~\cite{vLLM,sglang,flexgen,alisa}, fast GPU kernels~\cite{fastertransformer,flashattention,flashattention2,flashinfer}, and flexible decoding strategies~\cite{specinfer,sequoia}.
A truly system-optimal solution should ideally balance a wider array of practical metrics, including latency, memory, and even energy consumption.
Moreover, the definition of compute-optimality fails to capture the hardware heterogeneity.
Due to differences in memory hierarchy, interconnect bandwidth, and kernel-level scheduling, the same model may exhibit drastically different scaling behavior on various computing platforms~\cite{tpuv4,waferllm}.

In this work, we propose to investigate TTS from a system-driven perspective, where the focus is shifted from theoretical computation to practical system measurements.
Specifically, we aim to explore the scaling behavior of reasoning models to system metrics, namely, latency and cost-per-token, and discover the lack of system consideration of existing TTS.
Furthermore, we evaluate how prevalent optimization strategies, such as tensor parallelism and speculative decoding, would help achieve better system-oriented TTS.
Our goal is to conduct a preliminary analysis that provides a more comprehensive understanding of TTS's true performance.
Ultimately, we aim to call for a paradigm shift in how the AI and system communities evaluate reasoning strategies, moving beyond accuracy alone to embrace the holistic, system-level costs of intelligence.

\section{Related Work}
\minisection{Scaling Laws.}
Scaling laws first emerged as a powerful paradigm for neural networks in~\cite{scaling1}, where it shows that the language model performance, as measured by its test loss, improves predictably as a power-law with increases in model size (parameters), dataset size, and training compute. 
Their initial findings suggest that scaling model size is the most critical factor for achieving better performance. 
Later, Chinchilla conducts a more extensive analysis and concludes that for a given compute budget, optimal performance is achieved not by training the largest possible model, but by training a comparatively smaller model on significantly more data~\cite{scaling2}.
These principles have become an essential recipe in LLM pre-training, guiding the development of many subsequent popular state-of-the-art models~\cite{opt,cerebras,llama3,qwen2}.

\minisection{Test-time Scaling (TTS).}
Due to their ubiquitous performance on diverse language tasks, LLMs have become the de facto backend model executors for many popular applications~\cite{chatgpt,cluade}.
With the growing demands of real-world LLM-based applications, achieving proper scaling at inference time has become increasingly important.
Test-time scaling (TTS) methods aim to improve the LLM performance at relatively low costs on complex reasoning tasks such as mathematical reasoning~\cite{cot,s1}.
Notably, chain-of-thought (CoT)~\cite{cot} aims to leverage specialized instructions to guide pre-trained LLMs to generate better answers, and S1~\cite{s1} curates high-quality datasets and employs supervised fin-tuning (SFT) to achieve better TTS scaling.
Nonetheless, existing methods generally consider achieving compute-optimal instead of system-optimal.
Kinetics explores TTS scaling from sparse attention, but its system metrics are limited to per-layer latency~\cite{kinetics}.
Our work aims to reveal TTS scaling from a more comprehensive system angle. 

\minisection{LLM Systems.}
The proliferation of LLM-based applications has driven the development of numerous system-level designs to enhance inference efficiency. 
System advancements encompass kernel-level optimizations,~\cite{flashattention,flashattention2,flashinfer}, efficient memory management~\cite{vLLM,sglang,flexgen,alisa}, and flexible scheduling strategies~\cite{orca,fastserve,alise}.
Notably, the open-sourced vLLM~\cite{vLLM} and SGLang~\cite{sglang} have become the default serving engines for production-level system backends.
However, the workload characteristic of long sequence reasoning has not been fully understood in these systems. 
Our work aims to take the initial step in bridging the gap.

\section{Analysis}
\subsection{Experimental Setup}

\minisection{Models and Tasks.}
We choose two types of popular reasoning models for preliminary analysis, DeepSeek-R1-Distilled-Qwen~\cite{deepseekr1,qwen2} and S1.1~\cite{s1}, with three model size configurations, namely 1.5B, 7B, and 14B.
We use the MATH500 dataset~\cite{math500} and report the zero-shot accuracy results, and set the output length from 1K to 16K.

\minisection{System Setups.}
We use a private cluster with 4 NVIDIA GH200-96GB GPUs for evaluations, and build a serving engine using vLLM v.0.7.3 with PyTorch 2.6.0 and CUDA 12.6.
All evaluations are running using the FlashAttention v.2.7.2 backend.
For system-level metrics, we report both the average end-to-end latency per request and the cost per token measured as the number of GPUs used times latency, and divided by the total generated tokens.

\minisection{Optimization Strategies.}
To study the effect of existing optimization methods of system-level TTS, we select two popular methods, namely speculative decoding with a simple n-Gram predictor model~\cite{specinfer} and tensor parallelism~\cite{megatron}.
We set the number of speculative tokens to 5 and the minimum and maximum lookup to 1 and 4, respectively.

\begin{figure}[!ht]
     \centering
         \includegraphics[width=\linewidth]{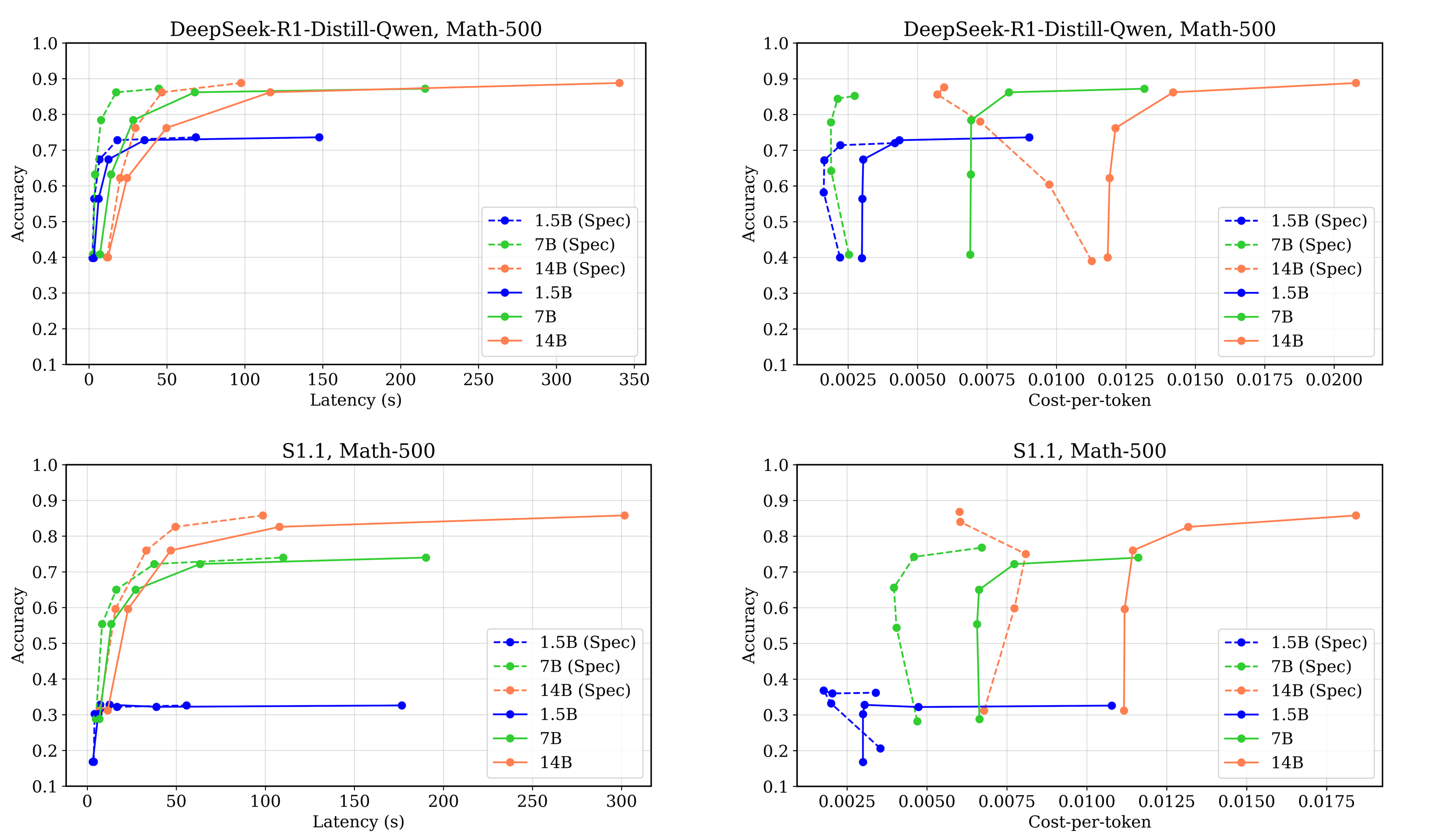}
        \caption{System-level scaling behavior comparison of different models applying speculative decoding on the MATH500 dataset.
        }
     \label{fig:main-1}
\end{figure}

\subsection{Results}

\begin{figure}[!t]
     \centering
         \includegraphics[width=\linewidth]{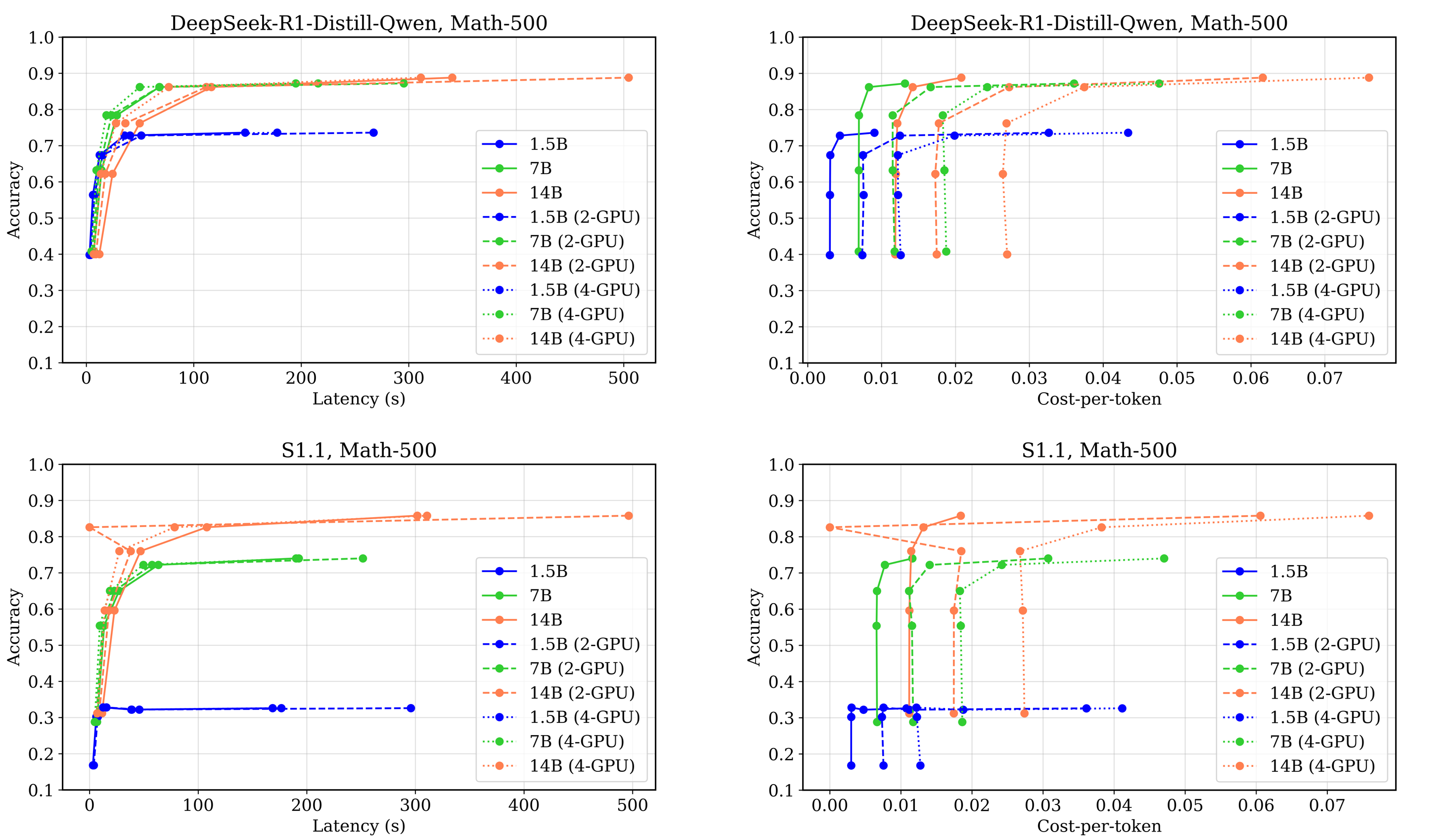}
        \caption{System-level scaling behavior comparison of different models applying tensor parallelism with different numbers of GPUs on the MATH500 dataset.
        }
     \label{fig:main-2}
\end{figure}

Figure~\ref{fig:main-1} and Figure~\ref{fig:main-2} demonstrates the preliminary results.
In terms of latency scaling, we have three key observations
First, the accuracy performance improves with more time spent on the reasoning process, but the gains gradually diminish and flatten out after a certain threshold.
This is in contrast to prior compute-oriented scaling, where the accuracy is linear with respect to the number of generated tokens.
Second, even with a simple N-gram model, speculative decoding has consistently shown improvement in latency over the naive greedy decoding strategy, revealing its great potential for application to larger models.
Third, naive scaling of the inference to more GPUs with tensor parallelism does not yield proportionally better results.
For instance, in the case of DeepSeek-R1-Distilled-Qwen-14B, scaling to 4 GPUs only improves the latency improvement by an average of 1.7$\times$;
for 1.5B models, multi-GPU results are even worse than single-GPU.
Such a phenomenon can be potentially attributed to the workload nature of reasoning tasks, emphasizing long sequences instead of large batch sizes.
Therefore, it makes the inference bottlenecked by the intra-GPU synchronization.

In regard to cost-per-token analysis, two key insights emerge from the results.
First, speculative decoding has shown better performance than tensor parallelism in reducing token costs, due to its latency speedup.
However, the scaling behavior is much more unpredictable than baseline methods in the case of larger 14B models.
This could be attributed to the randomness of the speculative model, where it cannot guarantee the same number of correct tokens at each step.
Second, due to the poor latency performance, tensor parallelism results in higher token costs than baseline methods.
This echoes the fact that compute-optimal does not equal system-optimal.

\section{Conclusion and Discussion}
In this work, we introduce a new system-oriented perspective on investigating test-time scaling (TTS).
Our key insight is that compute-optimal is not naturally equivalent to system-optimal, where the latter is often more important in real-world deployment with limited compute, memory, and energy.
Our results show the potential of speculative decoding and the limitations of tensor parallelism for system-oriented TTS.

While our work only shows results for limited workload scenarios, the underlying takeaway that reaching the true TTS Pareto frontier demands considerations of holistic system-level costs holds for different models and hardware.
By advocating for a shift from traditional accuracy vs. computation to system-aware evaluations, this work aims to provide a new angle to validate the practicality of existing methods in real-world scenarios.
This would encourage researchers to innovate on model architectures and inference strategies that are not just more accurate, but also resource-efficient, leading to a new class of LLMs optimized for low latency and cost.
Ultimately, this work calls for a more mature and pragmatic approach in developing TTS methods, which focuses on both the algorithm advancements and the corresponding suitable system designs.

\section{Acknowledgment}
\label{sec:ack}
This work was sponsored in part by the U.S. National Science Foundation (NSF) under Grants 2426368 and 2400014.
This work used DeltaAI at UIUC NCSA through allocation CIS250367 from the Advanced Cyberinfrastructure Coordination Ecosystem: Services \& Support (ACCESS) program, which is supported by U.S. NSF grants 2138259, 2138286, 2138307, 2137603, and 2138296.

\bibliographystyle{plain}
\bibliography{ref}

\end{document}